\documentclass[usenatbib,letterpaper]{mn2e}
\usepackage[totalwidth=480pt,totalheight=680pt]{geometry}
\usepackage{times,amssymb,epsfig,aas_macros}

\usepackage{color}
\definecolor{dr}{rgb}{0.6,0,0}
\definecolor{db}{rgb}{0,0,0.6}

\usepackage[ hyperfootnotes={false},
dvips,
pdftitle={Searching for Saturn's Irregular Satellite Swarm with Spitzer},
pdfauthor={Grant Kennedy},
pdfstartview={FitH},
bookmarksopen={true} ]{hyperref}
\hypersetup{colorlinks={true}, urlcolor={db}, citecolor={dr}}

\newcommand{\rs}{$R_{\rm Sat}$}

\begin{document}

\title[SEEKING SATURN'S SWARM]{Searching for Saturn's Dust Swarm: Limits on
  the size distribution of Irregular Satellites from km to micron sizes}

\author[G. M. Kennedy, M. C. Wyatt, K. Y. L. Su, J. A. Stansberry]
{G. M. Kennedy,$^1$\thanks{Email:
    \href{mailto:gkennedy@ast.cam.ac.uk}{gkennedy@ast.cam.ac.uk}} M. C. Wyatt,$^1$
  K. Y. L. Su,$^2$
  and J. A. Stansberry$^2$ \\
  $^1$Institute of Astronomy, University of Cambridge, Madingley Road, Cambridge CB3 0HA,
  UK\\
  $^2$Steward Observatory, University of Arizona, 933 North Cherry Avenue, Tucson, AZ
  85721, USA
}

\maketitle

\begin{abstract}
  We describe a search for dust created in collisions between the Saturnian irregular
  satellites using archival \emph{Spitzer} MIPS observations. Although we detected a
  degree scale Saturn-centric excess that might be attributed to an irregular satellite
  dust cloud, we attribute it to the far-field wings of the PSF due to nearby Saturn. The
  Spitzer PSF is poorly characterised at such radial distances, and we expect PSF
  characterisation to be the main issue for future observations that aim to detect such
  dust. The observations place an upper limit on the level of dust in the outer reaches
  of the Saturnian system, and constrain how the size distribution extrapolates from the
  smallest known (few km) size irregulars down to micron-size dust. Because the size
  distribution is indicative of the strength properties of irregulars, we show how our
  derived upper limit implies irregular satellite strengths more akin to comets than
  asteroids. This conclusion is consistent with their presumed capture from the outer
  regions of the Solar System.
\end{abstract}

\begin{keywords}
  Solar System: satellites
\end{keywords}

\section{Introduction}

The Solar System's irregular satellites are long thought to have undergone
collisions. When only eight were known at Jupiter, \citet{1981Icar...48...39K} showed
that the collision time for objects in the prograde group was less than the Solar
System's age. More recently, the discovery of collisional families shows that collisions
occurred in the past \citep{2003AJ....126..398N}. Further evidence lies with the
relatively flat size distributions of large irregulars, which \citet{2010AJ....139..994B}
show can be produced from initially steeper distributions by billions of years of
collisional evolution. Their collisional evolution is thought to have begun when the
irregulars were captured from cold icy regions of the Solar System
\citep[e.g.][]{2007AJ....133.1962N}.

Though the \citeauthor{2010AJ....139..994B} results are based on reproducing the size
distribution of the known irregulars, the destruction of the largest objects must produce
vast numbers of fragments. The fragments collide with each other, producing yet more
fragments and so on down to dust. The known irregulars are just the largest objects in a
continuous size distribution that extends down to the smallest grains that can survive on
circumplanetary orbits. It is therefore inevitable that dust associated with irregular
satellites exists at some level around the giant planets.

We recently made predictions of the level of this dust based on a collisional model and a
simple prescription for the size distribution \citep{2011MNRAS.412.2137K}. However, these
predictions are uncertain for several reasons. The many orders of magnitude between the
10--100km size of known objects and the smallest dust grains means that small differences
in the slope of the assumed size distribution result in large differences in the level of
dust. This slope depends on the strength law, which is uncertain. There may also be
additional loss mechanisms that flatten the size distribution relative to theoretical
predictions for a collisional cascade.

\begin{figure}
  \begin{center}
    \hspace{-0.35in} \psfig{width=0.47\textwidth,figure=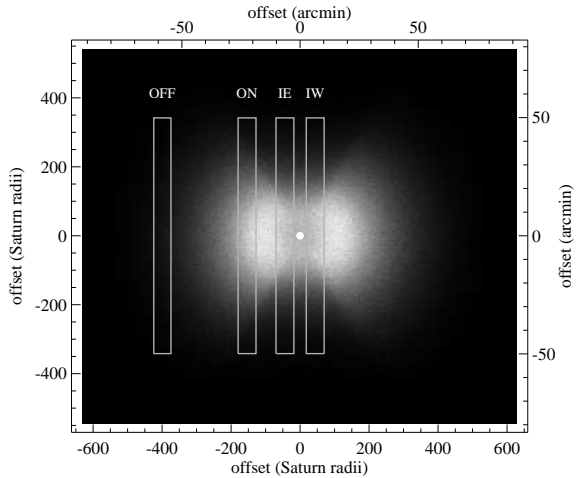}
    \caption{Simple model of Saturn's irregular satellite dust cloud in Saturnian and sky
      coordinates (y=0 is the ecliptic). The scale is a linear stretch. Saturn's position
      is marked by a dot (brightness not to scale). The boxes show the approximate
      position of the two Phoebe ring (ON,OFF) and two Iapetus (East,West)
      scans.}\label{fig:saturnpic}
  \end{center}
\end{figure}

Detection of this dust is therefore important for several reasons. Primarily it would
provide the strongest evidence yet that the known irregular satellites are just the tip
of a continuous size distribution. Because the collisional lifetime of the dust is of
order tens of millions of years, any existing dust must be continually replenished by
destruction of larger objects, which implies ongoing destruction of objects of all
sizes. Characterising the dust population is also important for understanding irregulars
themselves, because the slope of the size distribution depends on their strength
properties \citep{2003Icar..164..334O}. More generally, the irregular satellites provide
a rare chance to observe how the size distribution in a collisional cascade extrapolates
from large objects down to dust, thus informing models used to model extrasolar debris
disks.

Additional motivation comes from the existence of dust related to a single Saturnian
irregular satellite; the Phoebe ring \citep{2009Natur.461.1098V}. Impacts that launch
grains from Phoebe are proposed as a way to feed the ring. Whether the impactor
population is interplanetary, or belongs to a cloud of circumplanetary dust (i.e. due to
irregulars) is not known. Detection of a dust cloud would provide compelling evidence in
favour of the latter.

A simple way to estimate the spatial distribution of such a grain population is to assume
that they follow the orbits of the observed irregulars, which yields a model for the
surface brightness distribution of the dust cloud. Figure \ref{fig:saturnpic} shows this
model for Saturn, made by generating a population of particles with random semi-major
axes, eccentricities, and inclinations in the range derived for captured Saturn-centric
orbits by \citet{2007AJ....133.1962N}. The scale shows that the cloud extends over a
degree in each direction. The smaller vertical extent is the result of a lack of
near-polar orbits, which are unstable due to Solar perturbations
\citep{2002Icar..158..434C,2003AJ....126..398N}. The peak surface brightness predicted
for Figure \ref{fig:saturnpic} is 1.25MJy/sr \citep{2011MNRAS.412.2137K}, but as noted
above is uncertain.

Our aim here is to test this prediction for Saturn's dust cloud using the 24$\mu$m
\emph{Spitzer Space Telescope} observations used to discover the Phoebe ring. The regions
of sky covered by these observations relative to Saturn is also shown in Figure
\ref{fig:saturnpic}. The data cover a sufficiently wide region around Saturn to appear
promising for a deeper look for a cloud of dust originating from the irregular
satellites. In the next sections, we describe our efforts to extract a signal that can be
compared with Figure \ref{fig:saturnpic}.

\section{Data}\label{sec:data}

The \emph{Spitzer} data (programmes 40840 and 50780) are described in
\citet{2009Natur.461.1098V} and their location in relation to Saturn is shown in Figure
\ref{fig:saturnpic} \citep[see also][]{2009Natur.461.1098V}.\footnote{In Figure 1 of
  \citet{2009Natur.461.1098V} the Iapetus scans are rotated by 180$^\circ$, so their East
  image is actually the West one and vice versa. This mistake is presumably because
  \emph{Spitzer} would have been upside-down relative to the position for the ON/OFF
  scans.} For this study we use only the 24$\mu$m Multiband Imaging Photometer
\citep[MIPS,][]{2004ApJS..154...25R} data.

The Phoebe ring discovery data consist of a long image 128--180\,\rs\ from Saturn (called
ON, 18/2/2009), just inside the maximum extent of Phoebe's orbit. To allow foreground and
background subtraction, another image 400\,\rs\ away was also taken (OFF, 10/2/2009). In
addition, we found it necessary to use two images located much closer to Saturn (IAPETUS
East and West, 28/6/2008) to characterise the point spread function (PSF).

Each image comprises two consecutive scans 1.5$^\circ$ long taken approximately
perpendicular to the ecliptic. Each scan is built from 178 individual basic calibrated
data (BCD) images. The second scan is in the reverse direction to the first, but with a
small offset so each pair of scans overlap by about half their width ($\sim$2.5
arcmin). Here, we worked with the scans rather than the combined images (see
below). Before analysing the scans we remove point sources, Zodiacal foreground using the
COBE/DIRBE Zodiacal cloud model of \citet{1998ApJ...508...44K}, and mask a number of
glint and diffraction artefacts that are present due to the proximity of Saturn to the
field of view, particularly in the ON scan \citep[see][]{2009Natur.461.1098V}.

The zodiacal foreground subtraction does not account for all of the foreground in any of
the scans because ecliptic dust components \citep[e.g. asteroidal dust
bands;][]{1984ApJ...278L..19L,2001Icar..152..251G} are poorly characterised in this (or
any) model. However, because the ON and OFF scans were taken about a week apart, we only
expect an absolute offset due to the $\sim$9$^\circ$ change in elongation. We expect the
latitudinal structure to be similar, which is borne out by the data (see below).

Unfortunately, subtracting the OFF scan from the Iapetus scans cannot be justified due to
the seven month time difference, and the resultant different Zodiacal structure. Instead
the DIRBE model was scaled up by a factor 1.17 to bring the profiles to near-zero at the
ends. Alternatively, a DC offset can be subtracted to give essentially the same result.

\section{Analysis}

The data analysis is split into two sections. First, we do the ON-OFF subtraction, which
clearly shows excess surface brightness near Saturn. The two main possible causes of this
signal are i) our proposed dust cloud, and ii) the far-field wings of the Spitzer PSF due
to nearby Saturn (which is $\sim$15000Jy at 24$\mu$m, several thousand times greater than
the MIPS 24$\mu$m saturation limit of $\sim$6Jy). In the second subsection, we use the
Iapetus scans to characterise the PSF at large angular separations as a possible cause of
the measured ON-OFF signal.

\subsection{ON-OFF Subtraction}\label{sec:onoff}

Based on Figure \ref{fig:saturnpic}, we expect the cloud to vary smoothly over the
1.5$^\circ$ scan with a Gaussian-like shape. For the ON-OFF subtraction to be valid and
successful we are assuming that the dust cloud is not present in the OFF image, or is at
a much lower level. The dust level must also vary over the ON image, because some
constant offset is likely between the two images due to the elongation difference and
poor characterisation of ecliptic dust, which could not definitely be attributed to dust
associated with Saturn.

\begin{figure}
  \begin{center}
      \hspace{-0.35in} \psfig{width=0.52\textwidth,figure=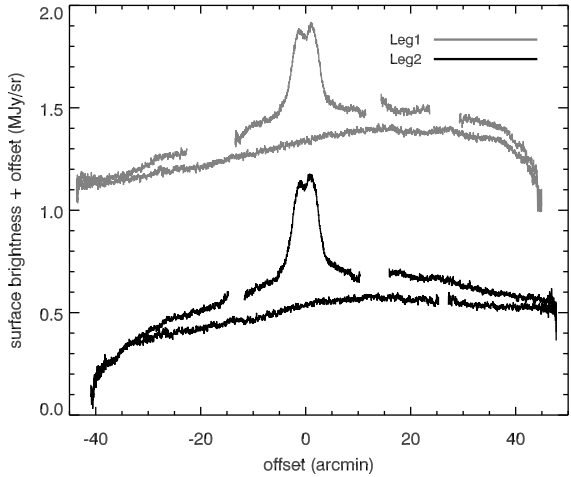}
      \caption{Zodiacal light subtracted median scan profiles. The x axis shows the
        approximate offset from the center of the Phoebe ring, visible as the enhancement
        in the ON scans. Gaps in the profiles are from masking out diffraction spikes and
        glints from Saturn. An arbitrary DC offset has been subtracted, with the Leg 1
        data offset for clarity.}\label{fig:raw}
  \end{center}
\end{figure}

Profiles across the ON and OFF scans are shown in Figure \ref{fig:raw}. Each profile is a
median collapse taken parallel to the long axis of a scan. Each pair has been scaled to
give the same level at the profile ends. The ON scans are easily identifiable by the
large bump in the middle due to the Phoebe ring. The corresponding leg of the OFF scan is
shown below each ON scan. Ecliptic South is to the left, so Leg 1 is taken from right to
left, and Leg 2 from left to right. The strong curvature at the beginning of each scan
(at offset of +40 arcmin for Leg 1, at -40 arcmin for Leg 2) is due to a brief change in
the detector bias applied just prior to the start of each scan. While the first five
frames are usually disregarded in typical scan maps, Figure \ref{fig:raw} shows that this
artefact is repeatable and can be accounted for as long as the individual scans are used.

\begin{figure}
  \begin{center}
    \psfig{width=0.47\textwidth,figure=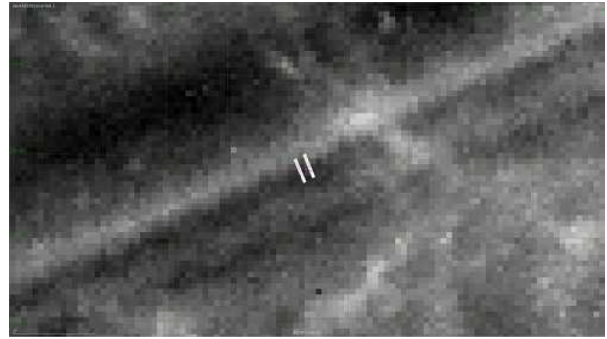}
    \caption{COBE DIRBE 25$\mu$m ZSMA image and MIPS ON/OFF scans (center, OFF is the
      left rectangle). The bright strip passing from lower left to upper right is
      parallel to the ecliptic, and shows excess surface brightness due to a poorly
      characterised heliocentric dust band.}\label{fig:zsma}
  \end{center}
\end{figure}

A positive trend towards ecliptic North (right) is apparent for all four scans, most
likely due to a heliocentric dust band. Whether such a trend is likely can be gauged
using Zodiacal model Subtracted Mission Averaged (ZSMA) COBE DIRBE maps. The ZSMA image
of the region observed in the ON/OFF observations is shown in Figure \ref{fig:zsma},
which makes it clear that the Zodiacal model is far from perfect in the ecliptic
\citep[see also Figure 2 of][]{1998ApJ...508...44K}. This image clearly shows residuals
that are the result of a heliocentric dust band not included or not well characterised by
the \citet{1998ApJ...508...44K} Zodiacal model. The most prominent band of excess
brightness coincides with the North end of the MIPS scans, which is likely the cause of
the observed increasing trend to the North end of the profiles (and perhaps some DC
offset). We cannot simply subtract a profile based on this image because the COBE DIRBE
maps comprise scans at a range of elongation angles.

\begin{figure}
  \begin{center}
     \hspace{-0.35in} \psfig{width=0.52\textwidth,figure=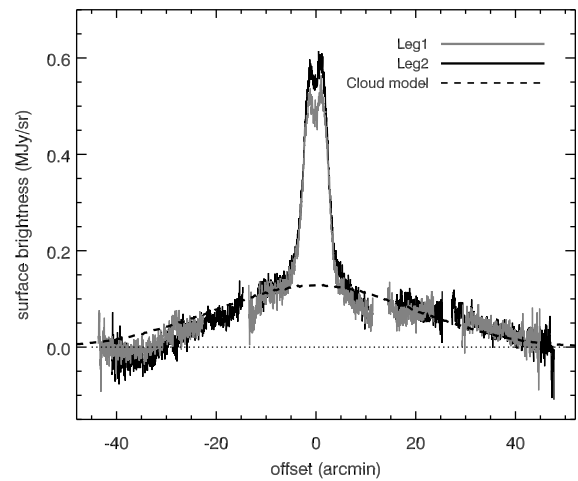}
  \caption{ON-OFF subtracted median scan profiles. The x axis shows the approximate
    offset from the center of the Phoebe ring.}\label{fig:sub}
  \end{center}
\end{figure}

As well as the obvious detection of the Phoebe ring, it is clear from Figure
\ref{fig:raw} that there is a difference that extends for nearly the entire length of the
ON and OFF scans. This difference is shown in Figure \ref{fig:sub}. This figure also
shows that the curvature due to the detector response at the start of each scan is
removed fairly successfully. Also shown is a vertical cut through the dust model of
Figure \ref{fig:saturnpic} at the position of the ON scan, scaled to match the observed
profile. The signal is clearly real, and tantalising as a possible dust detection.

\subsection{PSF Characterisation}\label{sec:iap}

While there appears to be a real difference between the ON and OFF images, we need to
consider the level of Spitzer's PSF due to Saturn, which could result in a similar
signal. The obvious way to estimate the level of the PSF is with the STinyTim PSF
model. However, these PSFs are well characterised on the scale of a few minutes of arc,
not the $\sim$degree scales that apply to these observations (J. Krist,
priv. comm.). Therefore, we must use an empirical approach, made possible by the
existence of the Iapetus scans. These scans are much closer to Saturn than the ON and OFF
ones, and may therefore be suitable for deriving an empirical profile of surface
brightness with radial distance from Saturn.

Of course the limitation of this approach is that the Iapetus scans may themselves have
detected the dust cloud, so consistency between the ON-OFF and Iapetus profiles does not
exclude the possibility of a dust cloud. Based on Figure \ref{fig:saturnpic}, one might
expect the level of dust to be lower in the Iapetus scans due to the non-spherical shape
of the cloud, in which case any significant difference between radial profiles extracted
from the Iapetus scans and the ON-OFF subtraction might be attributed to a dust cloud.

\begin{figure}
  \begin{center}
    \hspace{-0.35in} \psfig{width=0.52\textwidth,figure=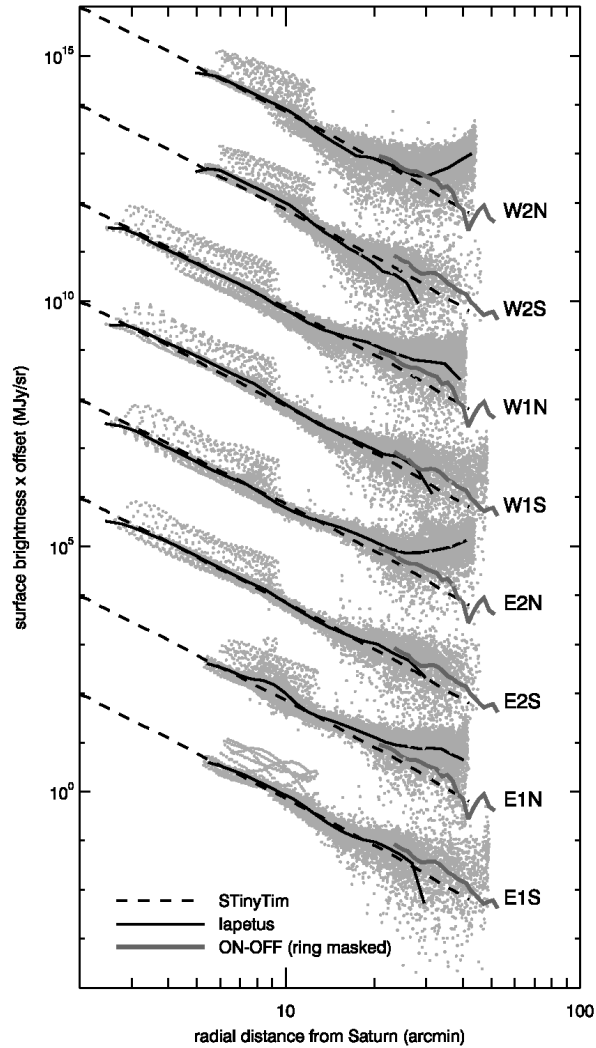}
    \caption{Median radial profiles derived from the Iapetus scans (dark solid lines) and
      STinyTim (dashed lines). Each scan is labelled according to whether it is East or
      West, Leg 1 or Leg 2, and North or South of a line along Saturn's orbit. The solid
      grey lines are median radial profiles of the ON-OFF scans, with Leg 1 plotted over
      the South Iapetus profiles and Leg 2 plotted over the North Iapetus profiles. The
      grey background dots are actual pixel values from the scans plotted at their radial
      distance from Saturn.}\label{fig:iap-log}
  \end{center}
\end{figure}

Figure \ref{fig:iap-log} shows median radial profiles of the two pairs of Iapetus scans,
centered on Saturn's position. The grey background dots are pixel values from the scans
plotted at their radial distance from Saturn. Taking the median means that azimuthal
structures such as diffraction spikes (visible at radii $\lesssim$10 arcmin) do not
affect the profiles. The profiles are also separated into North and South of a line along
Saturn's orbit (solid lines), and are labelled as `E' (East) and `W' (West), with the
scan legs following the same convention as the ON/OFF scans. Therefore the lowest profile
in the Figure, 'E1S', is the South part of Leg 1 of the Iapetus East scan.\footnote{Leg 2
  was actually taken first for the Iapetus observations (see first footnote)}

The reason for separating the scans into North and South is apparent. All North scan
profiles turn up around 30 arcmin from Saturn, likely due to the same dust band seen in
the ON/OFF scans in Figure \ref{fig:raw} but more pronounced. The difference is probably
because the Iapetus scans were taken seven months earlier with a consequently different
line of sight through the Zodiacal cloud, which unfortunately means we cannot use the OFF
scan to remove the foreground as we did for the ON scan. This turn up is less pronounced
in the Leg 1 (downward) scans (i.e. E1N and W1N), where the detector bias response (\S 2)
at the beginning of a scan counteracts the turn-up.

Figure \ref{fig:iap-log} also shows radial profiles derived from the ON-OFF subtracted
scans (assuming axisymmetry, grey lines). Despite uncertainties introduced by the dust
band, these profiles lie within the scatter of the Iapetus scans (i.e. sometimes above,
sometimes below), so appear consistent with having the same origin. The dashed lines show
a scaled STinyTim median radial profile, which also appear consistent with the Iapetus
observations. The STinyTim model is well fitted by a power law $\propto r^{-3.1}$ at
distances larger than 1 arcmin.

\begin{figure}
  \begin{center}
    \hspace{-0.35in} \psfig{width=0.52\textwidth,figure=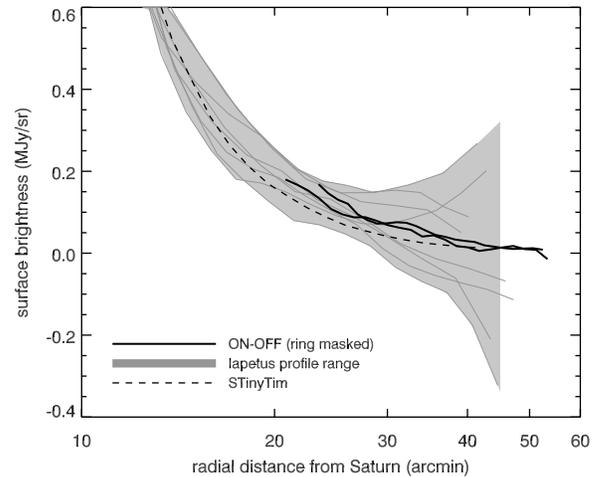}
    \caption{Median radial profiles derived from the Iapetus scans (grey swathe and dark
      grey lines) and STinyTim (dashed line). The median ON-OFF profiles (solid black
      lines) lie well within the swathe.}\label{fig:iap-lin}
  \end{center}
\end{figure}

Figure \ref{fig:iap-lin} again shows the radial profiles, but this time with a linear
y-axis. The shaded region is indicative of the uncertainty, and covers the range of
profiles for comparison with the ON-OFF and STinyTim profiles. The width of this swathe
beyond 30' is caused by the dust band (positive deviation) or detector bias response
(negative deviation). The $\sim$0.3MJy/sr negative deviation for the Leg 2 scans is
consistent with the deviation seen for the ON/OFF scans in Figure \ref{fig:raw}. The
STinyTim profile is bracketed by the empirical Iapetus profiles at all radii, suggesting
that the model is a good realisation of the \emph{Spitzer} response at large angular
scales. However, it is of course possible that the signal seen in the Iapetus scans is
partly or entirely due to our dust cloud and that the PSF contribution is negligible. The
near agreement between the Iapetus and ON-OFF profiles could then be because the dust
cloud is spherical and has less structure than Figure \ref{fig:saturnpic}. Unfortunately,
such conjecture cannot be tested with these data.

Given that the ON-OFF profiles are bracketed by the Iapetus profiles, the simplest
interpretation is that both have the same origin (i.e. the PSF). In this case an upper
limit on the emission from a dust cloud can be derived by subtracting the lower bound of
the Iapetus profiles from the ON-OFF profile. Doing so yields an upper limit about a
factor two lower than the profile shown in Figure \ref{fig:sub}. However, because we
cannot rule out the possibility that the PSF in fact has a negligible contribution, the
dust emission could be as much as in Figure \ref{fig:sub}. We adopt this more
conservative upper limit, and note the effect of the more stringent limit below.

\section{Discussion}\label{sec:disc}

There are two main points worth discussing; what the issues will be for future
observations that attempt to look for the same signal and whether they can be overcome,
and what we learn about irregular satellites from our upper limit on the surface
brightness of irregular satellite dust.

\subsection{Constraints on Irregular Satellites}

Although our characterisation of the Spitzer PSF suggests that the excess emission seen
in Figure \ref{fig:sub} is due to the wings of the PSF rather than a dust cloud, the
derived upper limit on dust sets interesting constraints on the size distribution of
irregular satellites.

The cloud model shown in Figure \ref{fig:sub} has a total flux of 45Jy. This flux is
derived by scaling the model in Figure \ref{fig:saturnpic} so that a profile at the ON
image position matches the level in Figure \ref{fig:sub}. To estimate the surface area
$\sigma_{\rm tot}$ in dust required to generate this level of thermal emission, we assume
blackbody grains at a temperature of $T= 88$K. Using the relation $F_\nu =
B_\nu(\lambda,T) \, \sigma_{\rm tot} / a_{\rm Sat}^2$ (where $a_{\rm Sat}=9.5$AU is
Saturn's semi-major axis), we therefore find an upper limit to the surface area in dust
of $\sigma_{\rm tot} = 1.3 \times 10^{-9}$AU$^2$ (or $2.9 \times 10^7$km$^2$). For
comparison, the projected area of the model in Figure \ref{fig:saturnpic} is about
0.15AU$^2$ so the limit on the optical depth is extremely low.

This limit assumes grains absorb and emit like blackbodies, but for more realistic grain
properties the surface density limit can only be more constraining. Real grains emit
inefficiently at wavelengths longer than their physical size, and are therefore hotter
than the equilibrium blackbody temperature. The difference here is minor because the
smallest grains are relatively large (16$\mu$m). Depending on the grain properties
(porosity, composition, crystallinity) the emission can be the same or a factor few
larger. For the highly porous grains inferred for comet-like dust
\citep[e.g.][]{1998A&A...331..291L}, the real grain emission is about a factor two larger
than for a blackbody. Because there are uncertainties in the grain properties and derived
spectra we retain the more conservative blackbody limit, and note the differences real
grains make below.

This upper limit can be converted to a limit on the average size distribution between the
smallest grains and largest objects, assuming a single phase size distribution that
follows $n(D) = K \, D^{2-3q}$ \citep[e.g.][]{2007ApJ...663..365W}, where $D$ is the
planetesimal diameter in km, and $K$ the normalisation. We assume the smallest grains are
of size $D_{\rm min} = 16\mu$m, approximately the smallest grains that can survive in
orbit around Saturn \citep{1979Icar...40....1B,2011MNRAS.412.2137K}. There are about 10
Saturnian irregular satellites larger than 10km in diameter, so the normalisation using
the cumulative number is $K = 10 (3q-3)/(10^{3-3q})$. Using the conversion from the size
distribution to total surface area $\sigma_{\rm tot}$ in grains
\citep{2007ApJ...663..365W}
\begin{equation}
  \sigma_{tot} = 3.5 \times 10^{-17} K (10^{-9} D_{\rm min})^{5-3q} / (3q-5)
\end{equation}
we can solve for the unknown $q$, which yields $q<1.81$. That is, if $q$ were larger, the
size distribution would be steeper and there would be more dust than our upper
limit. This maximum average size distribution index is similar to the commonly used
canonical value of $q=11/6$, derived for a collisional cascade size distribution where
strength is independent of size \citep{1969JGR....74.2531D}. The upper limit on
$\sigma_{\rm tot}$ is three times lower than would have been predicted with $q=11/6$.

\begin{figure}
  \begin{center}
    \hspace{-0.35in} \psfig{width=0.52\textwidth,figure=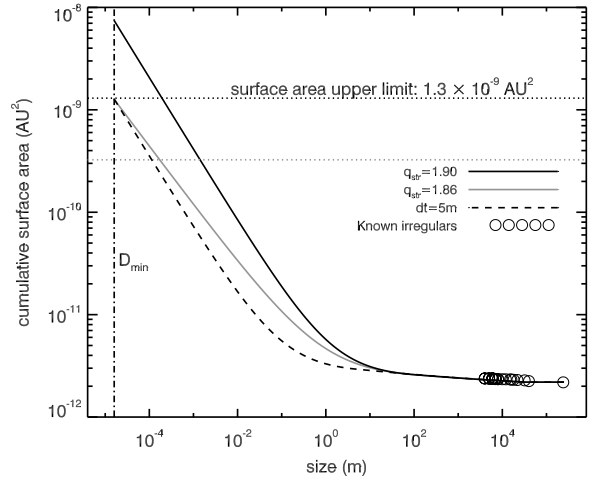}
    \caption{Cumulative surface area plot showing theoretical distributions and the known
      irregular satellites of Saturn. The black dotted line is the upper limit on the
      surface area implied by the MIPS observations. The grey dotted line is the less
      conservative upper limit.}\label{fig:sizedist}
  \end{center}
\end{figure}

Our predictions in \citet{2011MNRAS.412.2137K} were based on a two phase size
distribution, which fit the size distribution of known irregulars. The two phases arise
because object strength sets the size distribution slope \citep{2003Icar..164..334O}, and
there are two different strength regimes
\citep[e.g.][]{1998Icar..135..431D,1999Icar..142....5B,2009ApJ...691L.133S}. Above the
$\sim$0.1km transition size, objects gaining strength from self-gravity have a shallower
slope ($q_{\rm grav}$) than smaller objects, whose strength is derived from material
properties (and have a slope $q_{\rm str}$). The slope in the strength regime is
generally negative because larger objects are more likely to have a significant flaw
\citep[e.g.][]{1994Icar..107...98B}. The slope $q_{\rm str}$ is flatter (less negative)
for more porous objects \citep[e.g.][]{1990Icar...84..226H,2011Icar..211..856H}.

While studies generally agree that the strength and gravity regimes exist, the transition
size and dependence of strength on size varies between authors. In predicting a level of
320Jy for Saturn's dust cloud \citet{2011MNRAS.412.2137K}, we used a transition size of
0.1km, $q_{\rm str}=1.9$, and $q_{\rm grav}=1.7$.

Figure \ref{fig:sizedist} compares our upper limit with the prediction and two other size
distributions. All size distributions end at the assumed minimum grain size $D_{\rm
  min}=16\mu$m. The black solid line shows our prediction and the dashed line shows the
same distribution with a smaller transition size of 5m. The solid grey line shows the
same distribution with a slightly flatter size distribution in the strength regime
($q_{\rm str}=1.86$). The observed Saturn irregulars sit at the lower right, illustrating
the many orders of magnitude in size between the largest and smallest objects, and why
predictions are uncertain. We also include a less conservative (but not unreasonable)
limit on surface density, assuming that the Iapetus scans contain no dust and that the
dust is made of ``real'' comet-like grains (see above), which is a factor four lower.

Figure \ref{fig:sizedist} shows that there are several ways the surface area can be lower
than the upper limit. The surface area decreases as the minimum grain size $D_{\rm min}$
increases, and for our original prediction would need to be about a factor of ten higher
($\sim$160$\mu$m) to be consistent with the upper limit. Models that explore
circumplanetary grain dynamics in this context are therefore needed. If the effective
minimum grain size is larger our upper limit becomes less constraining, but if grains are
typically smaller it is more constraining.

Alternatively, the transition between the strength and gravity regimes may be smaller
than the assumed 0.1km. For the same strength dependence, the transition size needs to be
20 times smaller (5m) to satisfy our upper limit (dashed line). Though there are modest
variations, no models predict the transition size to be this small
\citep[e.g.][]{1999Icar..142....5B}.

A more likely possibility is that the size distribution in the strength regime is
flatter, as illustrated by the grey line. The slope of $q_{\rm str}=1.86$ corresponds to
a strength dependence of $\propto D^{-0.17}$. For the less conservative limit $q_{\rm
  str} = 1.83$ and the strength dependence is $\propto D^{0.06}$ (i.e. slightly
positive). This dependence is flatter than most theoretical models, which lie in the
range -0.2 to -0.6
\citep[e.g.][]{1990Icar...83..156D,1999Icar..142....5B,2009ApJ...691L.133S}. It is
significantly flatter than the strength dependence inferred for small asteroids impacting
Gaspra and Ida \citep[-1, ][]{1994Icar..107...84G,1996Icar..120..106G}. A likely reason
the strength dependence is flatter than typically predicted or observed is that small
irregulars are more porous than asteroids
\citep[e.g.][]{1990Icar...84..226H,2011Icar..211..856H}, and therefore more comet-like
\citep[e.g.][]{1990ApJ...361..260G,2002aste.conf..485B}. Strength properties more akin to
comets than asteroids are consistent with the proposed scattering and subsequent capture
or irregulars from the outer reaches of the Solar System
\citep[e.g.][]{2007AJ....133.1962N}. The difference between out upper limit of 45Jy and
the prediction of 320Jy may therefore be because the strength of small irregular
satellites depends less strongly on their size than we assumed, and the size distribution
consequently flatter.

However, the strength prescription could be correct because any additional loss processes
(other than collisional evolution) would lead to an overall flatter size
distribution. The slightly positive strength dependence for our less conservative upper
limit suggests that other loss processes do indeed occur. For Saturn, we estimated that
Poynting-Robertson (PR) drag is roughly as important as collisional evolution for
removing the smallest grains, and therefore possibly an important loss process that may
modify the small end of the size distribution
\citep{2011MNRAS.412.2137K,2011arXiv1103.5499W}. For the originally assumed size
distribution in Figure \ref{fig:sizedist}, the surface area would need to turn over
around 160$\mu$m, which is significantly larger than the minimum grain size and therefore
unlikely. However, as for the minimum grain size, grain dynamics need to be modelled in
detail to study the effect of radiation forces, including other possible effects
(e.g. Yarkovsky force).

Observationally, pushing the minimum known sizes of irregular satellites down will better
characterise their size distribution over a wider range. However, progress will be
difficult. Table 3 in \citet{2007ARA&A..45..261J} shows that the faintest known
irregulars at Saturn have R magnitudes of 24.5,\footnote{The faintest published survey
  has $R=24.5$ \citep{2001Natur.412..163G}, but \citet{2007ARA&A..45..261J} list an
  unpublished survey with a limiting magnitude of $R=26$.} and the deepest survey for
Uranian irregulars had a 50\% detection efficiency at $R=26$mag
\citep{2005AJ....129..518S}. Such an improvement for Saturnian satellites would push the
smallest objects down by a factor of about 4, to about 1km.

Complementary methods that probe smaller sizes are therefore also needed, such as crater
counts. While craters have been counted on Saturnian satellites
\citep[e.g.][]{1982Sci...215..504S,2010Icar..206..485K}, to back out a size distribution
of irregulars is complex. For example, because small and large impactors alter the
target's surface in different ways, the observed crater size distribution does not
necessarily reflect the impactor distribution
\citep[e.g.][]{1994Icar..107...84G,1996Icar..120..106G,2009Icar..204..697R}. Few studies
consider irregular satellites as a source of impactors
\citep[see][]{2010AJ....139..994B}, probably because most irregulars were only discovered
in the last ten years or so and it was only recently proposed that the known irregulars
are the tip of an iceberg of collisional fragments
\citep{2010AJ....139..994B,2011MNRAS.412.2137K}. \citet{2010AJ....139..994B} make a
simple comparison based on their irregular satellite evolution model, but also
acknowledge the complications of trying to derive a crater population from a population
of irregulars whose size distribution and dynamics evolve as they grind down. Given that
the number of irregulars was likely much greater in the past, they may have dominated the
impactor population and should be considered when modelling crater counts.

\subsection{Future observations}

In trying to extract the signal of our proposed dust cloud, we had to consider a number
of possible confounding issues. For the ON and OFF scans the Zodiacal contribution could
largely be accounted for because the Zodiacal contribution is expected to change little
with small changes in elongation. An alternative way to account for ecliptic dust it is
to take two observations separated by exactly one year. The first includes the region
around the planet, and the second observes the same patch of sky (the planet having moved
on in its orbit), and thus has the same line of sight through the ecliptic and the same
(extra)galactic background. However, the limiting factor for this study was in fact
Spitzer's PSF. PSF characterisation may be the most important issue for future
observations of this type.

Given that instrumental PSFs may be hard to avoid and characterise, an additional way to
maximise the chance of a detection is to look for dust around a different planet. The
argument for doing so could be due to either a better chance of detection (assuming a
similar dust cloud is present), or that the dust cloud is likely to be brighter. For the
former case, the giant planets are near their equilibrium temperatures so their spectra
are roughly similar to that expected for dust. Therefore, a contrast advantage can be
gained by looking at Uranus, which is much fainter and whose dust cloud is expected be of
similar surface brightness to that predicted for Saturn \citep{2011MNRAS.412.2137K}. The
question of which planet is more likely to harbour detectable dust is uncertain. Uranus
may again be more favourable than Saturn because the dust may be less affected by PR drag
\citep{2011MNRAS.412.2137K}. Dust from irregular satellites at Neptune is probably at a
lower level due to the proposed disruption of the population by Triton and/or Nereid
\citep{2003AJ....126..398N,2005ApJ...626L.113C}.

While our model considers a large uniform dust cloud, there is also the possibility of
smaller scale structures due to individual objects such as the Phoebe ring
\citep{2009Natur.461.1098V}. Because they are non-axisymmetric, these structures should
be easier to detect. The best candidates appear to be Phoebe, Nereid, and perhaps the
Carme collisional family, because their low inclinations relative to our line of sight
increase the surface brightness of any associated dust.\footnote{A face-on Phoebe ring
  would be about 15 times fainter or about 0.03MJy/sr. Such a faint ring would not have
  been detected in the Spitzer observations.}

\section{Summary}

We attempted to detect the Saturnian irregular satellite dust cloud proposed to exist by
\citet{2011MNRAS.412.2137K} using existing archival \emph{Spitzer} MIPS observations. A
signal was observed, but is consistent with being the PSF due to nearby Saturn, which is
likely to be the main issue for future observations. We therefore consider the detection
to be an upper limit on the level of dust in the outer reaches of the Saturnian
system. In the absence of other loss processes, the upper limit constrains the size
distribution of irregulars and therefore their strength properties, which we find to be
more akin to comets than asteroids. This conclusion is consistent with their presumed
capture from the outer regions of the Solar System.

\vspace{0.5cm}

This work is based in part on observations made with the Spitzer Space Telescope, which
is operated by the Jet Propulsion Laboratory, California Institute of Technology under a
contract with NASA. This research also made use of Tiny Tim/Spitzer, developed by John Krist
for the Spitzer Science Center. The Center is managed by the California Institute of
Technology under a contract with NASA. We thank John Krist for his advice on the
limitations of the STinyTim PSFs.

\bibliography{../ref} \bibliographystyle{astroads}

\end{document}